\documentclass[9pt,twocolumn,twoside]{optica}
\usepackage{graphicx}
\usepackage{caption}
\usepackage{amsmath}
\usepackage{amssymb}
\usepackage{color}
\usepackage{epstopdf}
\usepackage{authblk}
\setboolean{shortarticle}{false}
\setboolean{minireview}{false}

\title{Quantum receiver for large alphabet communication}

\author[1,2]{I. A. Burenkov}
\author[3]{O. V. Tikhonova}
\author[2,*]{S. V. Polyakov}

\affil[1]{Joint Quantum Institute \& University of Maryland, College Park, MD 20742, USA}
\affil[2]{National Institute of Standards and Technology, Gaithersburg, MD 20899, USA}
\affil[3]{M.V. Lomonosov Moscow State University, Moscow, 119991, Russia}

\affil[*]{Corresponding author: sergey.polyakov@nist.gov}

\ociscodes{(060.1660) Coherent communications; (270.1670) Coherent optical effects; (270.5585) Quantum information and processing; (060.2630)   Frequency modulation.}

\begin{abstract}
Quantum mechanics allows measurements that surpass the fundamental sensitivity limits of classical methods. 
To benefit from the quantum advantage in a practical setting, the receiver should  use communication channels resources optimally; this can be done employing large communication alphabets. 
Here we show the fundamental sensitivity potential of a quantum receiver for coherent communication with frequency shift keying.
We introduce an adaptive quantum protocol for this receiver, show that its sensitivity outperforms other receivers for alphabet sizes above 4 and scales favorably, whereas quantum receivers explored to date suffer from degraded sensitivity with the alphabet size. 
In addition, we show that the quantum measurement advantage allows much better use of the frequency space in comparison to classical frequency keying protocols and orthogonal frequency division multiplexing.
\end{abstract}

\setboolean{displaycopyright}{false}

\begin{document}

\maketitle

\section{Introduction}

Long-distance communication with light dates to at least 1184 BCE, when a series of fire beacons were used to signal the fall of Troy over 600 kilometers \cite{HomerClassic}. Over the following three millenia light continued to be used for long-distance communications, with the methods dramatically improving. Now, transfer rates of terabits/second are possible with modest powers  
of just a few thousand photons per bit incident at the receiver \cite{Maher2016}. The new paradigm of data storage and processing via virtualization and cloud computing \cite{Armbrust} motivates the exploration of  
novel telecommunication protocols \cite{Mecozzi,MarinPalomo}, particularly those that provide higher data transfer rates \cite{Maher2016,Bozinovic} and/or use channel resources more efficiently \cite{Millar,Maher15}. With the continuous proliferation of high data rate applications as a driver, 
the data rate of modern communication systems doubles nearly every 18 months \cite{Cherry}. Thus, the total energy required to transmit one bit becomes a fundamental factor hindering the development of networks. 
In current fiber optical channels maximum input power is limited due to nonlinear effects and the maximum distance between repeaters is limited to $\approx$100 km due to optical losses \cite{Agrell}. 
On the other hand, deep space communication systems are constrained  by the very limited power available on a spacecraft, making data rate improvement extremely difficult with traditional protocols \cite{Cesarone,Kaushal}.  
Coherent states  are currently the information carriers of choice. 
These states are naturally resilient to losses and can reliably carry information through amplitude, phase, and/or frequency modulation \cite{Proakis}. However, the accurate discrimination of these states is limited by inherent noise \cite{MandelBook}.
  Classical optical receivers 
are now approaching this ultimate sensitivity limit  known as the standard quantum limit (SQL) \cite{Robertson,Caves,Caves2,Kikuchi}. 
Quantum measurements can outperform their classical counterparts in sensitivity (energy required to obtain a given discrimination accuracy) and lead to important new applications in classical and quantum communications, quantum information processing, biophotonics, etc. \cite{Braunstein,Zoller,Taylor}. A quantum-enhanced receiver can exceed the restrictions of the SQL and push sensitivity toward a much lower fundamental error bound - the Helstrom bound (HB) \cite{Helstrom}. State discrimination below the SQL has been demonstrated in recent experiments with quantum receivers \cite{sasaki10,sasaki11}. 

\begin{figure*}
\centering
\includegraphics[width=0.7\columnwidth]{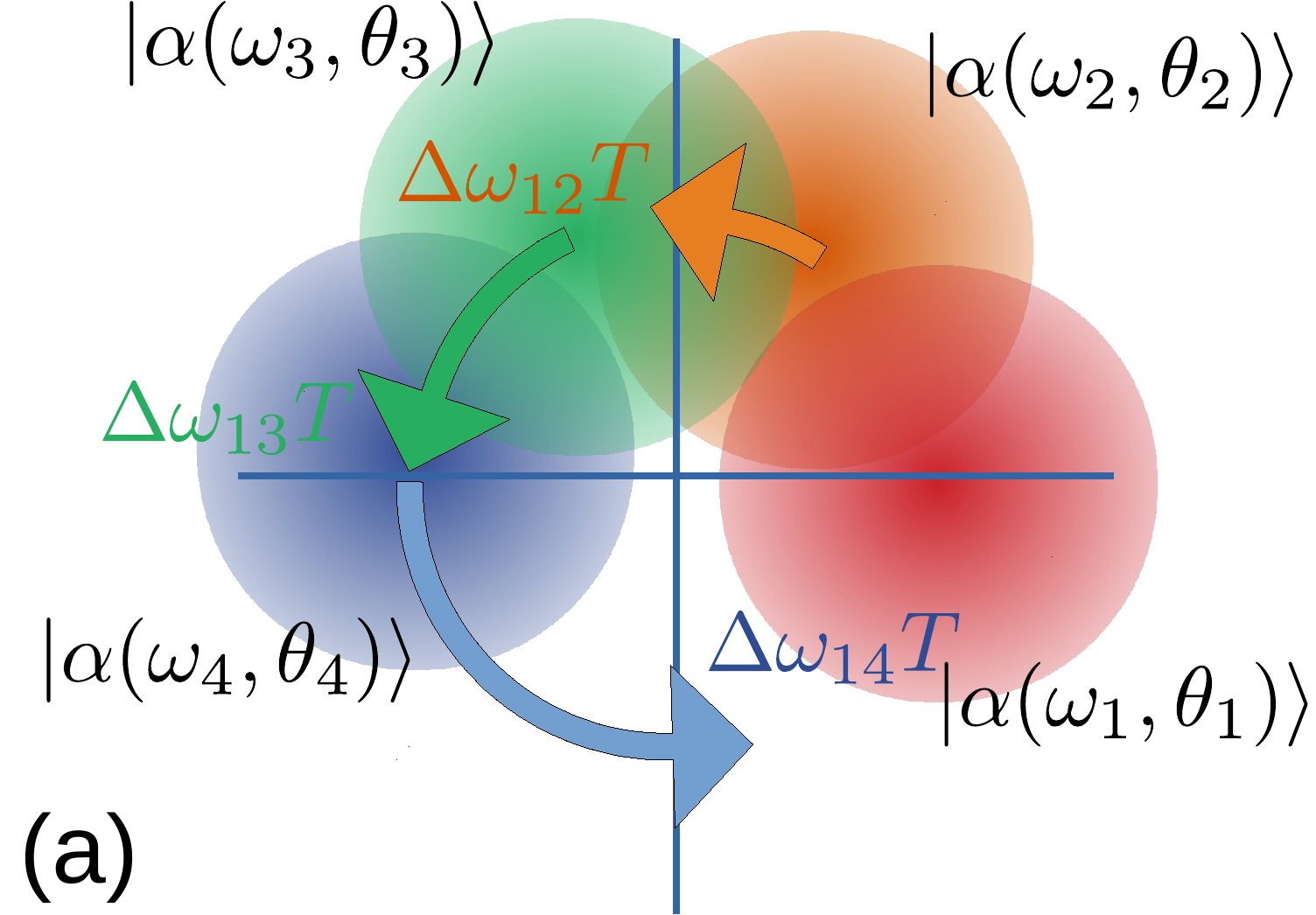}
\includegraphics[width=0.85\columnwidth]{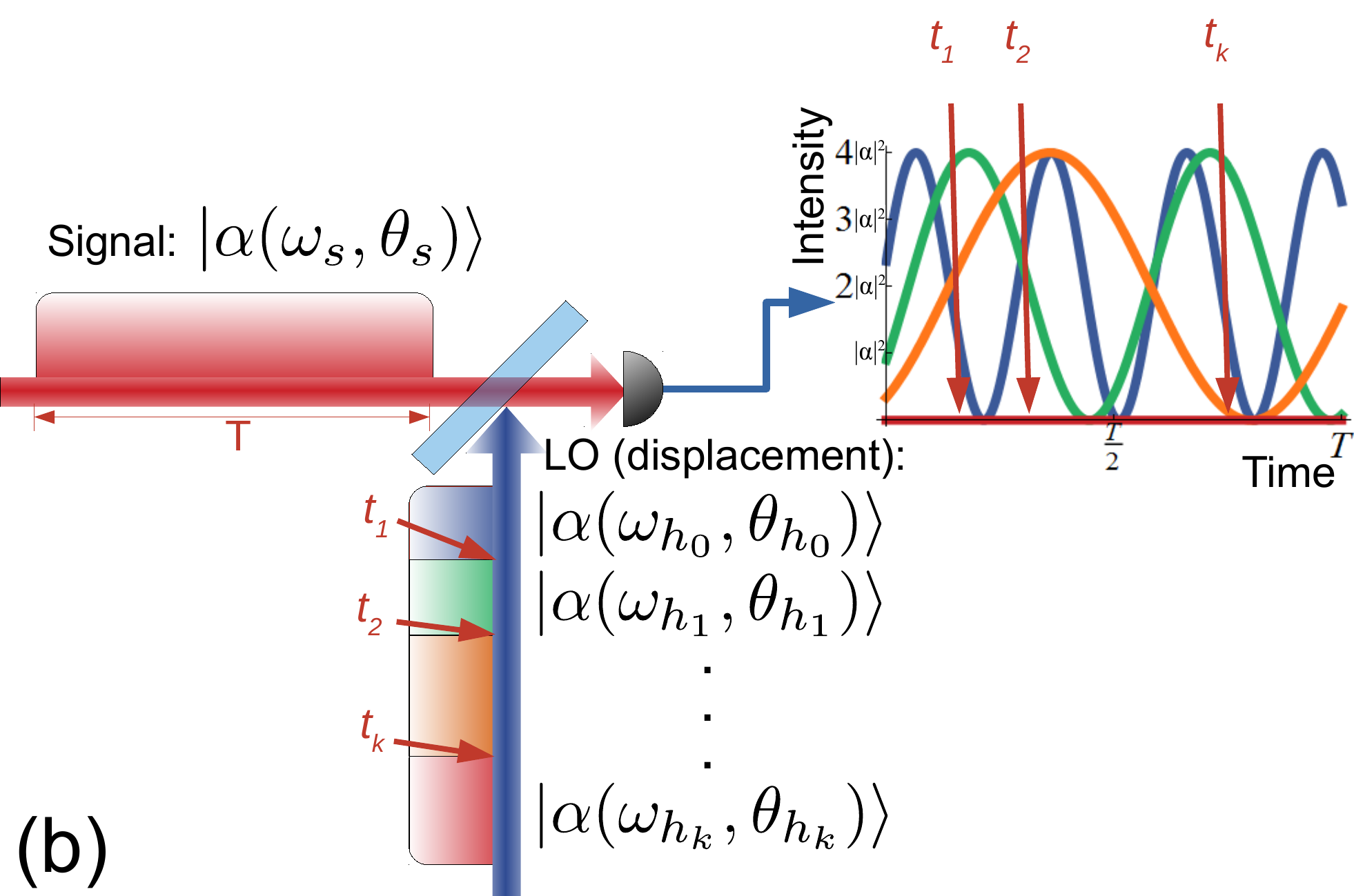}
\caption{\label{fig:method} (a) Constellation diagram introducing the $M$-ary coherent frequency shift keying (CFSK). Symbols are encoded as coherent states of different frequencies $\omega_i$ and phases $\theta _i$, resulting in rotation with time around the origin with rates that correspond to their detuning from the carrier frequency (shown in red).
(b) Quantum receiver's principle of operation. A displacement corresponding to the most probable state of the input field $|\alpha(\omega_s, \theta_s)\rangle$ is followed with a single-photon detector. A $k^{\mathrm{th}}$ click on the detector invalidates the hypothesis $h_k$. The probability of a click varies with time, hence the time of detection $t_k$ contains important information about the input state (inset). To find the best {\it a posteriori} hypothesis, we implement a Bayesian adaptive strategy for a continuous, time-resolved measurement, see text. 
}
\end{figure*}

Modern communication protocols have evolved to use large alphabets consisting of up to a few thousand symbols \cite{Koizumi,Chen}, which significantly improves the transfer rate and spectral capacity of a communication channel.
 There have been quite a few theoretical studies of quantum receivers for binary and larger alphabets. The protocols studied are  pulse amplitude modulation, where the information is encoded as the amplitude of a carrier wave \cite{Tsujino}, phase shift keying (PSK), where the information is encoded in phase \cite{Becerra2011,Mller}, and quadrature amplitude modulation (QAM), where a combination of amplitude and phase is used for encoding \cite{Zuo}. To date, quantum receivers that discriminate as many as 4 coherent states with error rates below the SQL have been experimentally demonstrated \cite{Becerra2013,Becerra2015}. However, most of heretofore explored quantum receivers suffer from sensitivity degradation with alphabet size. 


Here we introduce an $M$-ary quantum receiver based on coherent frequency shift keying (CFSK). CFSK offers an HB that is below the HBs of other receivers for large alphabets in a head-to-head comparison, i.e. with a fixed energy per encoded bit. In addition, our receiver exhibits much better sensitivity scalability with alphabet size, comparable only to pulse position modulation (PPM). However, unlike the PPM \cite{Takeoka}, our keying scheme is naturally immune to communication rate reduction with alphabet size.
Our receiver operates with a classical transmitter, and with any communication channel, including the existing global fiber network. Its advantages can be used to increase the distance between repeaters in a network and/or to reduce power requirements on the transmitter by more than 30 dB. In addition, the quantum measurement advantage can significantly optimize the use of frequency space in comparison to classical frequency keying, continuous phase modulation protocols, and orthogonal frequency division multiplexing (OFDM) \cite{Proakis}.

\section{Results and discussion}

\subsection {Quantum theory of CFSK: the protocol, error bounds, and scalability}

The CFSK protocol encodes information in the frequency and phase of coherent state pulses of duration $T$ using an alphabet of $M$ symbols. Symbols $m$ and $j$ are separated by $\Delta\omega_{m,j}$ in frequency space, and have initial phases $\theta_m$, $\theta_j$. This alphabet can be pictured using the constellation diagram, Fig. \ref{fig:method}(a). Coherent states corresponding to CFSK symbols rotate with time around the origin of the diagram with rates given by their detuning.

To facilitate the theoretical description of the protocol, we assume that signal pulses have rectangular shape. This assumption does not limit the generality of the proposed scheme, because other pulse shapes could be taken into account straightforwardly with an additional weighting factor. We consider equal separation between the adjacent symbols in frequency $\Delta\omega=\omega_{m+1}-\omega_m$ and phase $\Delta\theta=\theta_{m+1}-\theta_m$. 
Note that the pulse duration $T>0$ always appears in a product with $\Delta\omega$. 
Thus, the protocol is described by two independent parameters: $\Delta\theta$ and $\Delta\omega T$. Note that small detunings, $\Delta\omega T<2\pi$, are of particular interest because we aim at minimizing the bandwidth usage. 
This parameter space contains the PSK modulation scheme: $\Delta\omega T=0$, $\Delta\theta  =2 \pi / M$. The performance characteristics of PSK protocols are well-known \cite{Dolinar,Kennedy,Becerra2011} and will be used here as a benchmark for comparison. A fair comparison of energy efficiency between different encoding types and different sizes of the alphabet $M$ is a very important practical question.  We consider the encoding capacity of the alphabet, $\log_2 M$ in bits per symbol (BPS), rather than the alphabet size $M$  and express the discrimination error probabilities as a symbol error rate (SER).

We derive the SQL and HB for our detection strategy. In a classical receiver, errors occur when noise on the received input results in it being better correlated to a state other than the sent state. 
The SQL defines the lowest error bound for an ideal classical receiver due to shot noise of the input coherent states \cite{Caves,Jaekel}. 
To find the SQL we calculate error probabilities for spectrally overlapping coherent signals in the shot-noise limit. 

The HB is the minimal possible error probability for quantum discrimination of non-orthogonal states \cite{Helstrom}.
To obtain the HB, we use the square root measure method \cite{Belavkin,Holevo,Wootters,Kato}, see supplementary materials for a detailed derivation. 
The HB strongly depends on the modulation parameters $\Delta\omega T$, $\Delta\theta$  as shown in the supplementary material. This dependence can be used for optimization of the protocol.

\begin{figure}
\centering
\includegraphics[width=0.3\textwidth]{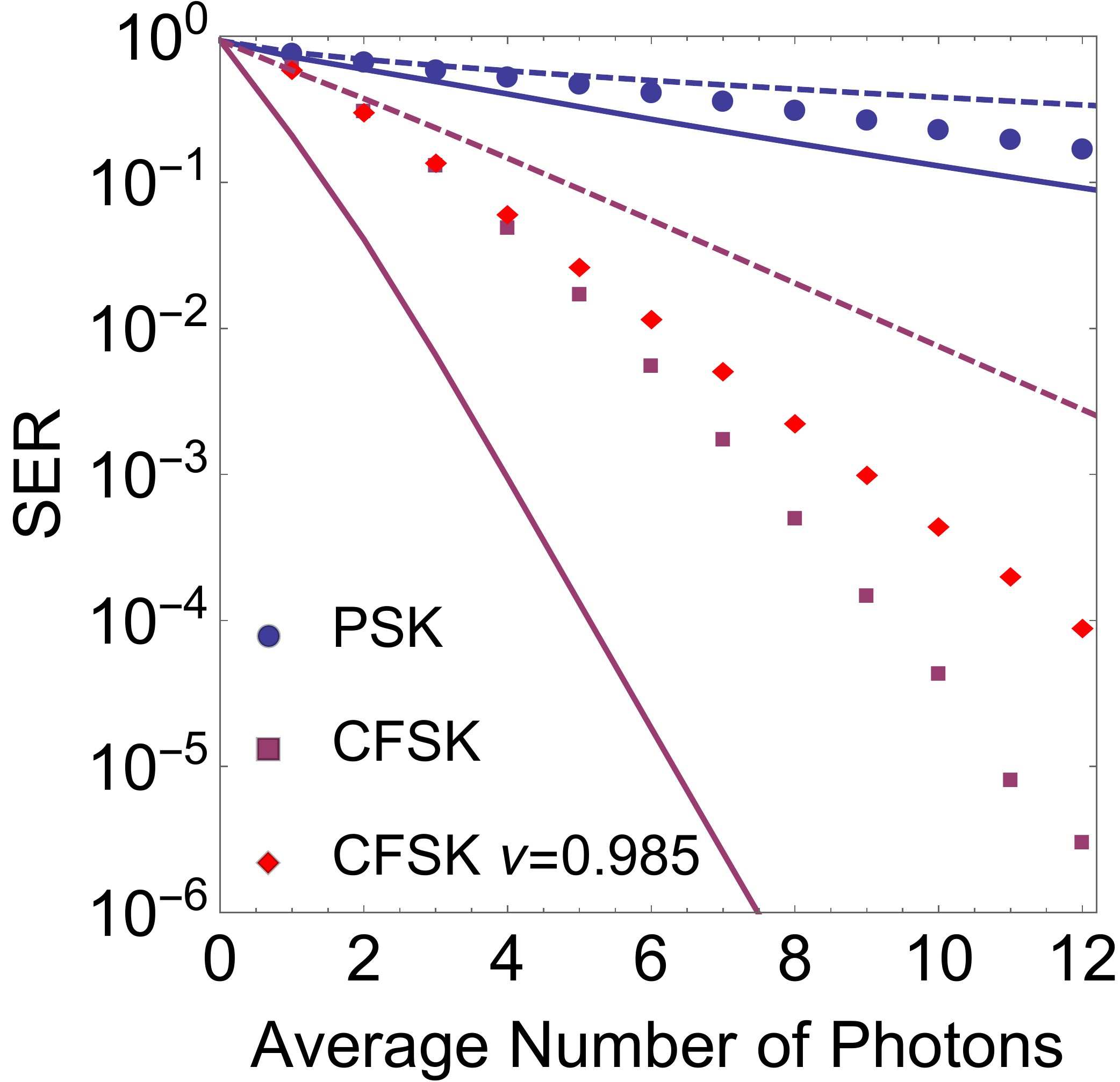}

\caption{\label{fig:power} Energy dependence of symbol error rates along with fundamental bounds for different encoding techniques with the 4-bit ($M=16$) alphabet. CFSK receiver: purple squares, CFSK receiver with practical visibility: red diamonds, PSK receiver: blue dots. Discrimination error bounds are shown with lines: PSK (blue) and CFSK (purple). SQL - dashed lines, HB - solid lines.}
\end{figure}

There is a sizable range of the parameters that yields much smaller HBs than possible for $M$-ary PSK. We optimize our receiver over these parameters (see supplementary material) and use the optimal parameter set for further analysis. Fig. \ref{fig:power} shows the dependence of the HBs and SQLs on the average number of photons in a pulse for an $M=16$ alphabet with encoding capacity of 4 BPS. Points represent results of numerical simulations for quantum receivers and are described below.
In addition to a well-researched PSK protocol, we consider the performance of the lowest-order QAM protocol with a square constellation diagram (e.g. \cite{Proakis}) and a PPM protocol \cite{Dolinar} for the same alphabet length, which covers all 4-bit ($M=16$) quantum receiver schemes proposed to date. 
We find that the CFSK HB is below HBs of other data transfer protocols for any input energy at the receiver. 

\begin{figure}
\centering
\includegraphics[height=0.29\textwidth]{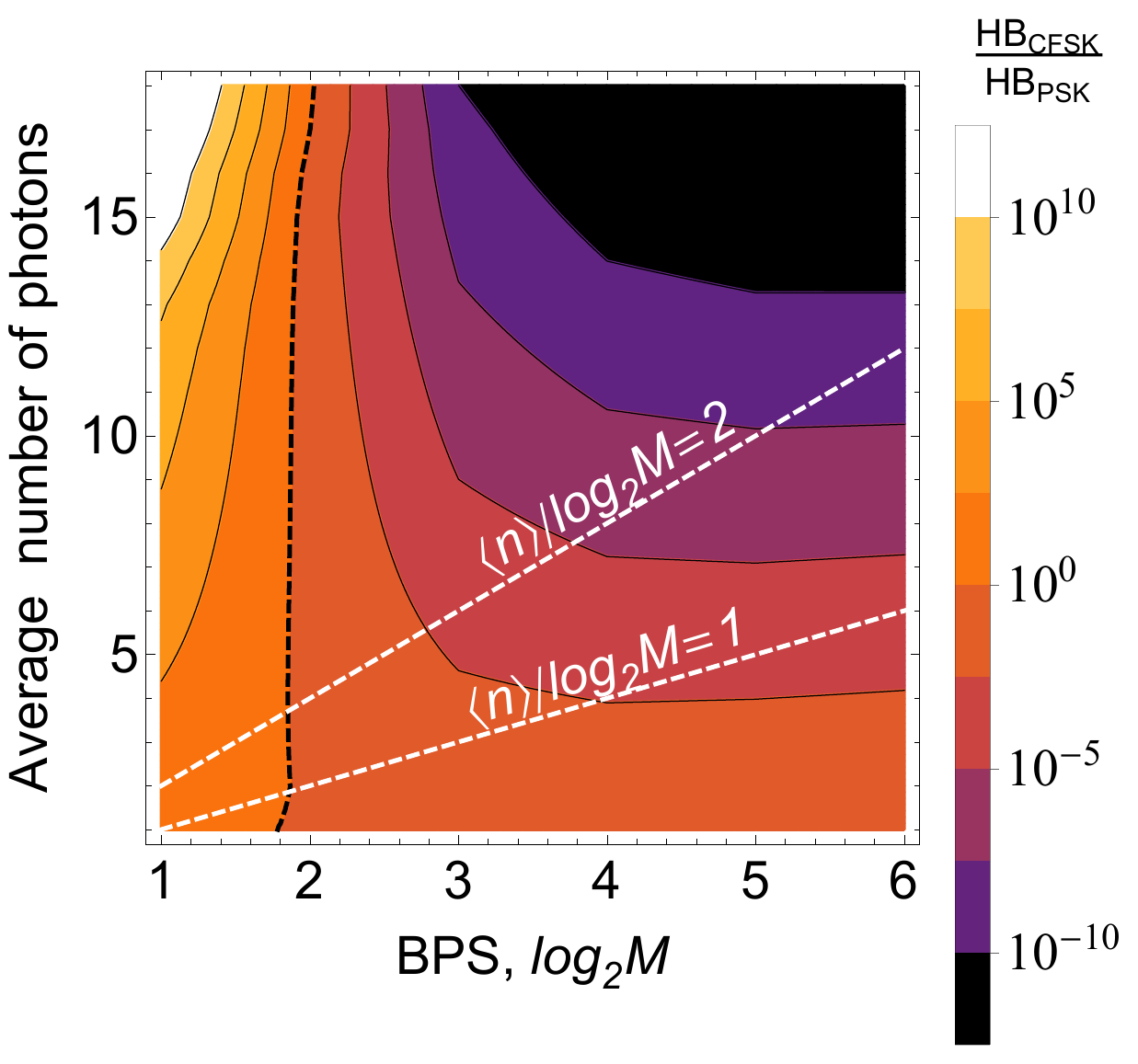}

\caption{\label{fig:biterrorrate1} Energy efficiency scaling. A ratio of HBs of CFSK and PSK over a range of input energies and alphabet lengths. A significant advantage of $M$-ary CPSK over $M$-ary PSK in terms of error probability is evident for all input signal energies for all $M>4$ alphabets, $\log_2M>2$. The region of CFSK advantage is all the area to the right of the black dashed line which indicates a HBs' ratio of 1. White lines correspond to constant input energy per bit to aid in comparison (see text).}
\end{figure}

Next, we establish scalability of CFSK with respect to alphabet size. We numerically calculated the HB for a range of input energies and alphabet sizes. The energy efficiency of CFSK improves with the alphabet size, whereas the energy efficiency of PSK rapidly decreases with the alphabet size. This data is best presented as the ratio of HBs for CFSK and PSK, Fig. \ref{fig:biterrorrate1}. We see that CFSK outperforms PSK for $M>4$ at any input energy (area to the right of the black dashed line in Fig. 3). The observed advantage grows with the alphabet length and input energy. We note that the primary reason to employ quantum receivers is to improve  energy efficiency for the data transfer. 

\subsection {Temporally resolved CFSK quantum receiver}

A CFSK receiver harnesses the quantum advantage through the use of adaptive displacement. The strategy tests if the hypothesized state $|\beta_k \rangle = |\alpha(\omega_{h_k}, \theta_{h_k})\rangle$ corresponds to the state of the input field $|\psi\rangle =|\alpha(\omega_s, \theta_s)\rangle$ by displacing the input state  by $|\beta_k \rangle$ on a 99:1 interferometer, Fig. \ref{fig:method}(b). Then, the displaced state $|\psi-\beta_k \rangle$ is measured with a photon-counting detector. If the hypothesis is correct ($\left|\beta_k\right\rangle=\left|\psi\right\rangle$), the input field is displaced to the vacuum and no photons are detected at the output. Because the states differ in frequency, the displacement of the input based on the incorrect hypothesis results in a field whose intensity depends on time differently for each combination of signal  $|\psi\rangle$ and LO $|\beta_k \rangle$, inset in Fig. \ref{fig:method}(b). In classical communication, the state can be recognized by Fourier transforming the intensity beating \cite{Mecozzi}. For a quantum receiver, if even a single photon is detected, the hypothesis fails. Still, photon arrival times bear additional information about the state of the input. Because single-photon detectors provide an accurate time stamp of a photon detection, we take advantage of the photon arrival times in calculating the \textit{a posteriori} most probable state $|\beta_{k+1} \rangle$ to update the hypothesis.
To derive this time-resolved Baesyan strategy, we find instantaneous average photon numbers in the displaced fields for all symbols of the alphabet and calculate the contribution of interarrival times to the posterior probability.

The light measured at the detector is a coherent state, whose mean photon number evolves in time $\langle n(m,h,t)\rangle$, where $m$ and $h$ denote the states of the receiver input and the LO used for displacement, respectively. This time dependence influences photon interarrival times measured at the detector, therefore providing information about the input state.
Photon interarrival times for a coherent state are governed by an exponential distribution depending only on mean photon number \cite{ross2007}. 
Therefore, when a photon arrives at time $t_k$,  the {\it a posteriori } probability that the input state equals $m$ is given by:
\begin{equation}
\zeta_{t_k}(m)=\frac{(\langle n(m,h,t_k)\rangle/T) e^{-\int_{t_{k-1}}^{t_{k}}\langle n(m,h,\tau)\rangle d\tau /T}\zeta_{t_{k-1}}(m)}{\sum^M_{j=1}(\langle n(j,h,t_{k})\rangle/T) e^{-\int_{t_{k-1}}^{t_{k}}\langle n(j,h,\tau)\rangle d\tau /T}\zeta_{t_{k-1}}(j)},
\end{equation}
where $t_{k-1}$ is the detection time of the previous photon, $\zeta_{t_{k-1}}(m)$ is the {\it a priori } probability that the input is in state $m$. For the first photon detection ($k=1$) the {\it a priori } probability is uniform for all states, $\zeta_0=1/M$.
This is a general recurrent expression for any modulation method, i.e. where the mean photon number at the detector can change arbitrarily (cf. \cite{Mller}). 

In the case of CFSK, the instantaneous mean photon number at the detector is given by $\langle n(m,h,t)\rangle=2\langle n \rangle \left(1-\cos \left[(h-m)(\Delta \omega  t+\Delta\theta  )\right]\right)$. The adaptive sequence of displacements is obtained recursively from the history of previous displacements and measured photon arrival times, i.e. every time a photon arrives at the detector, the hypothesis about the input state is updated to that with the highest {\it a posteriori} probability and the displacement is set correspondingly. 

\begin{figure}
\centering
\includegraphics[width=0.35\textwidth]{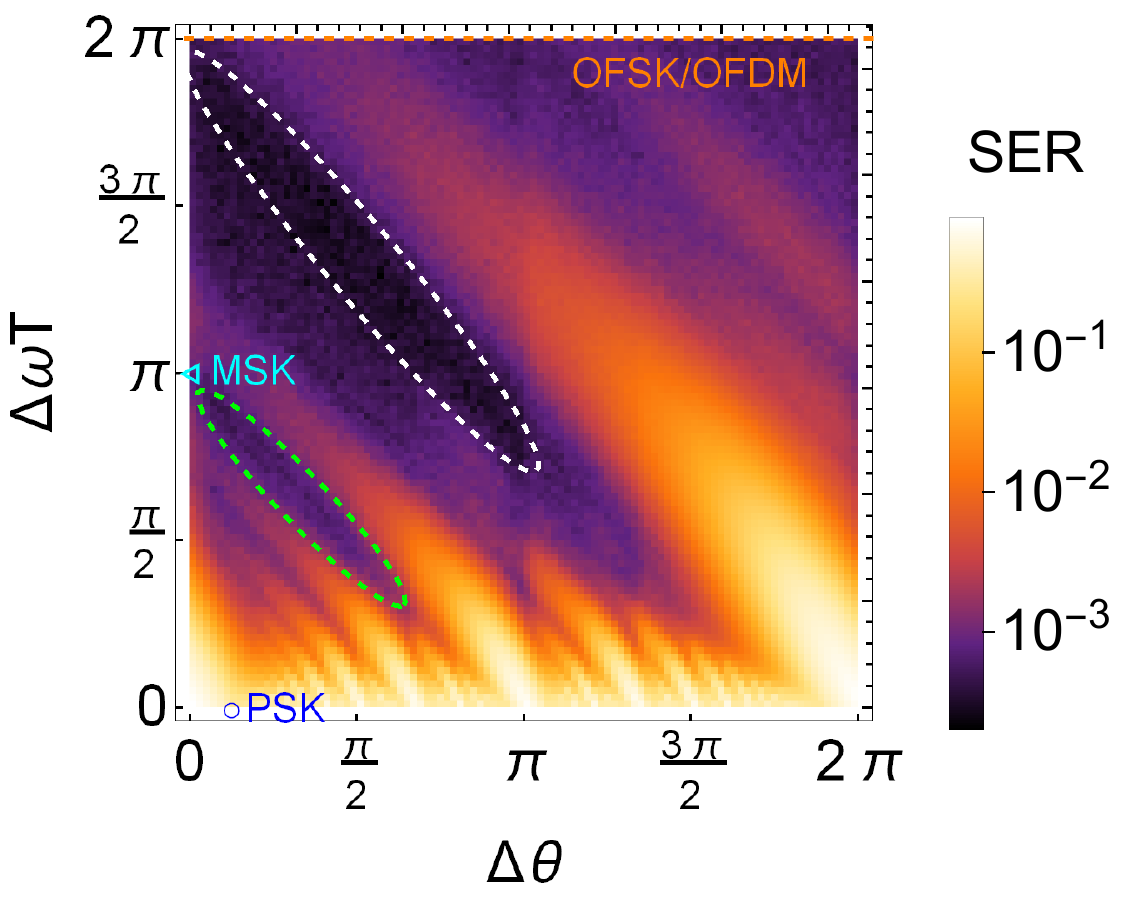}

\caption{\label{fig:plots3d} A Log symbol error rate (SER) map versus the optimization parameters for a CFSK receiver with a 4-bit ($M=16$) alphabet with the input energy of 2 photons per bit. The optimal parameter range for energy sensitivity: white dashed contour. The secondary optimal parameter range for energy sensitivity with a modest increase of symbol error rate, but with twice smaller frequency separation between the states $\Delta\omega T$, green dashed contour. The optimization parameter space contains the set that corresponds to PSK (blue circle). Our receiver beats the PSK receiver in SER by over three orders of magnitude. Parameter sets for the classical orthogonal FSK and OFDM (dashed orange line) and the minimum shift keying (light blue triangle) are marked to aid with comparison. }
\end{figure}

\begin{figure}
\centering
\includegraphics[width=0.35\textwidth]{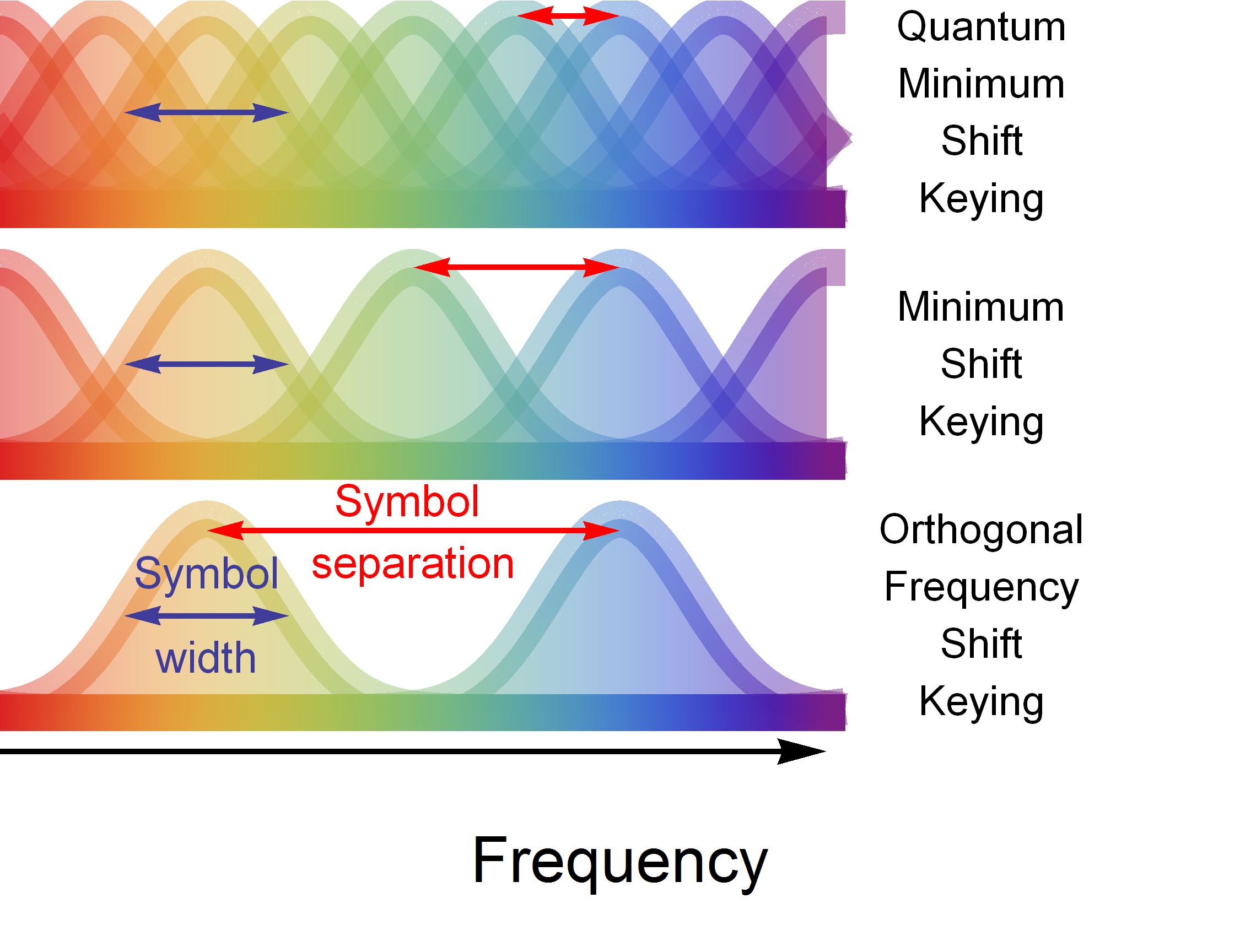}
\caption{\label{fig:spectral} A schematic comparison of spectral separation of classical vs. quantum frequency-shift-based receivers.}
\end{figure}

To characterize the quantum receiver we developed a Monte-Carlo-based numerical algorithm that simulates the adaptive protocol and calculates the SER for an ideal single-photon detector using the fraction of trials with an incorrect symbol discrimination result. 
Initially, we obtain optimization maps over $\Delta\theta $ and $\Delta\omega T$. The 4-bit ($M$=16) alphabet map is shown in Fig. \ref{fig:plots3d}. There are a few notable properties of this map. 
First, the parameter range that minimizes the SER (marked with the white dashed contour) is quite different from the orthogonality conditions used in classical receivers \cite{Proakis}. 
Indicated are the frequency separation employed in a classical \textit{minimum}  shift keying (light blue triangle labeled (MSK) and the parameter range used in classical orthogonal FSK and OFDM (orange dashed line).
Second, our receiver has a secondary minimum, whose vicinity is marked with a green dashed contour. Employing this minimum significantly improves the use of frequency space for the FSK with a slightly reduced energy efficiency. This regime gives an approximately four-fold advantage in bandwidth in comparison to an orthogonal FSK and OFDM and a nearly two-fold advantage in comparison with the MSK, see Fig. \ref{fig:spectral}. 

Next, we study the CFSK receiver performance dependence on energy and compare it to other encoding schemes,  Fig. \ref{fig:power}. Here we use the CFSK parameters that minimize the SER. 
We see that the SER advantage of the CFSK receiver over PSK receiver scales exponentially with energy. In addition, the CFSK receiver offers accuracy below the HBs of PSK and QAM, therefore, it establishes a fundamentally unreachable sensitivity for these receivers. 
Specifically, we find that the CFSK receiver yields an advantage of 42 dB over the PSK HB 
for a 4-bit (M=16) alphabet with average number of photons per symbol $\langle n \rangle=12$. Further, the CFSK receiver beats the  most efficient time-resolved \cite{Mller} (rather than multi-stage) PSK receiver by 45 dB under the same conditions. 

The CFSK advantage remains in place under realistic experimental conditions. Here we consider imperfect interference of the received signal with the local oscillator. We see that the error probability increases only slightly (to about 15 dB for 4-bit ($M=16$) alphabet and $\langle n\rangle=12$) for a reasonably attainable interference visibility of 98.5\% (c.f. \cite{Becerra2013}, where the interference visibility of 99.7\% was experimentally observed). The detection inefficiency also increases the SER. However, as it follows from this computation, a modest detection efficiency of about 70\% is sufficient to surpass the SQL of CFSK even with an imperfect visibility of 98.5\%, and the SER  remains well below the HBs of other protocols. 

\begin{figure}
\centering
\includegraphics[height=0.23\textwidth]{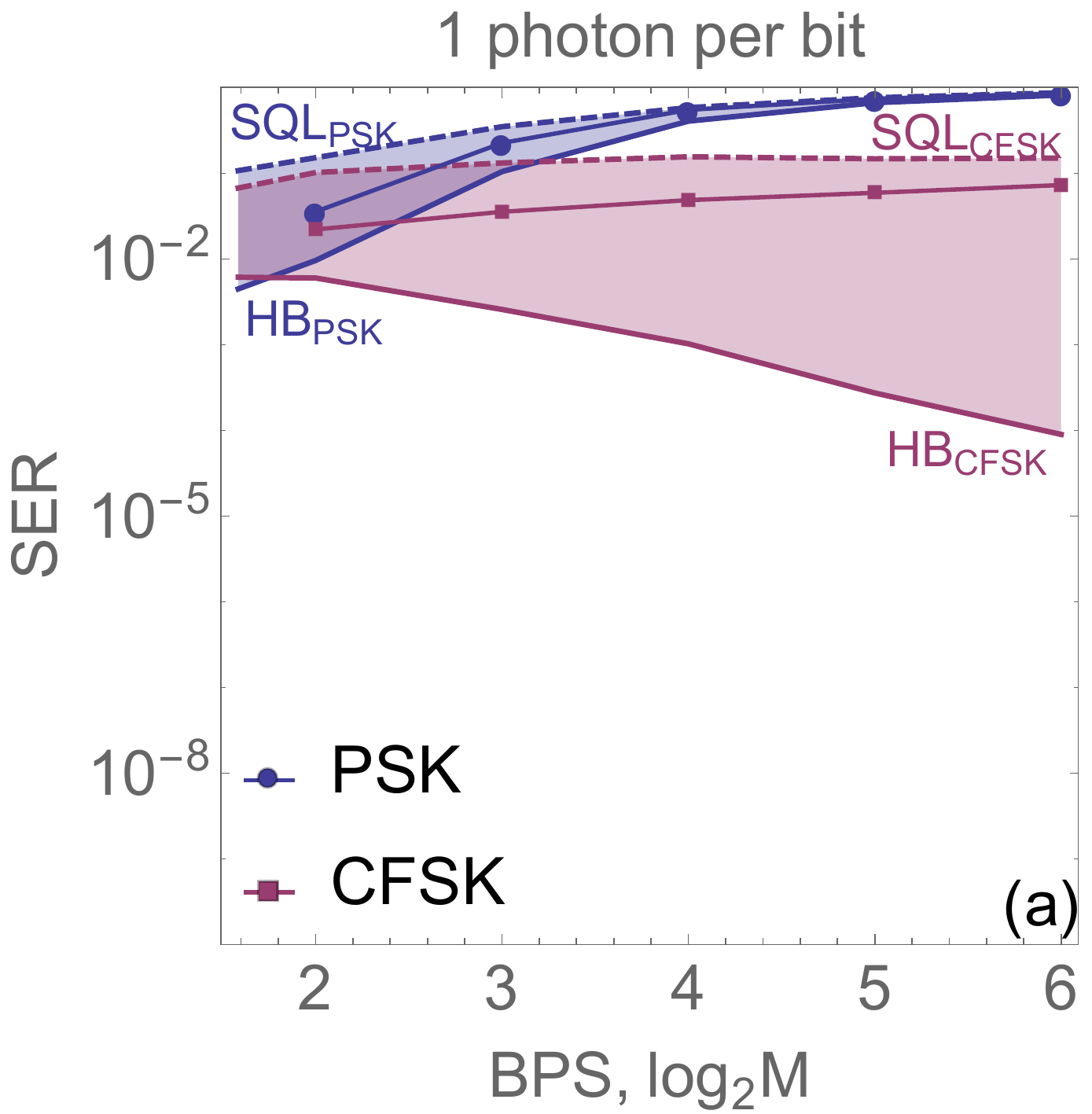}
\includegraphics[height=0.23\textwidth]{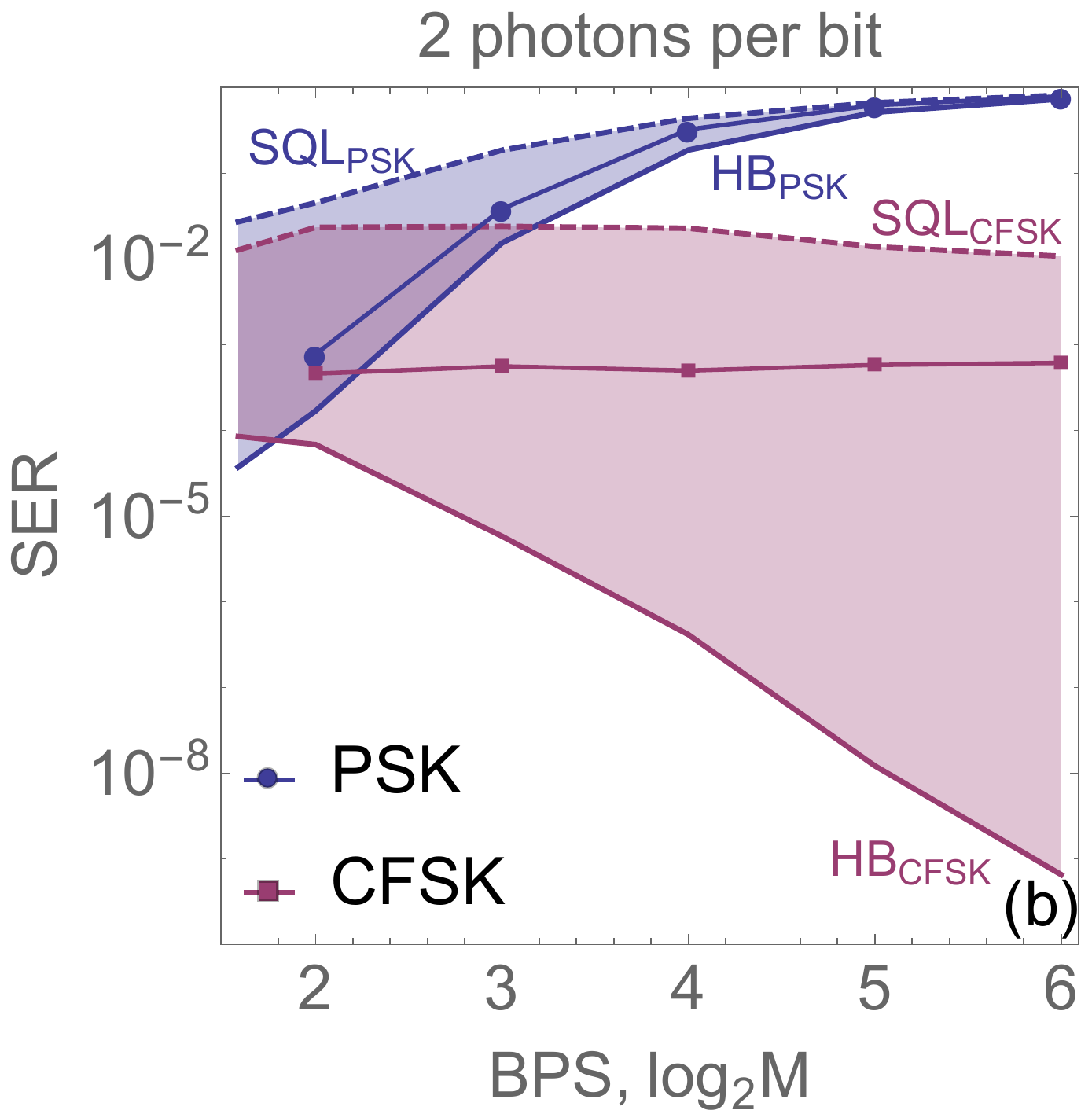}
\caption{\label{fig:biterrorrate2} Symbol error rate performance of quantum PSK (blue dots) and CFSK (purple squares) receivers for different encoding capacities $\log_2 M$ with a constant input energy per bit, cf. white lines in Fig. \ref{fig:biterrorrate1}. In addition, SQLs (dashed lines) and HBs (solid lines) are shown. The symbol error rate of a quantum PSK receiver significantly worsens with the longer alphabets, while that of our CFSK receiver provides scaling with the alphabet size. Thin lines connecting dots, squares are guides for an eye.}
\end{figure}

\subsection {Discussion}

To put the results of this work in perspective, we present theoretical bounds of the CFSK and the performance of this receiver together with that for the PSK for a range of alphabet lengths $M$, Fig. \ref{fig:biterrorrate2}. In doing so, we fix energy per transmitted bit $\langle n\rangle/\log_2 M={1,2}$ (white dashed lines in Fig. \ref{fig:biterrorrate1}). As it can be inferred from Fig. \ref{fig:biterrorrate2}, the quantum advantage of a PSK scheme decreases and one can show that the PSK SER always saturates to unity with alphabet length for any finite input energy.
In contrast, even at one photon per bit ($\langle n\rangle/\log_2 M=1$) CFSK demonstrates much better scalability with encoding capacity $\log_2 M$, Fig. \ref{fig:biterrorrate2}(a). For two photons per bit input the SER remains nearly constant for any encoding capacity of the alphabet, Fig. \ref{fig:biterrorrate2}(b). For even brighter coherent states ($\langle n\rangle/\log_2 M>2$) the SER decreases with alphabet length. 

This scalability of our quantum receiver enables its immediate use with large-sized alphabets that optimize the channel capacity. 
Note that the HB of the CFSK decreases even more rapidly with the encoding capacity of the alphabet. Closing the gap between our quantum receiver and the theoretical SER bound is therefore an important goal for the future research. 

Further, the secondary SER minimum found on the optimization map may be employed to significantly reduce the frequency band of a communication channel. In other words, this minimum can be used to develop a novel quantum minimum-shift keying protocol (QMSK). 
Because classical MSK-based protocols are so widely used in present-day digital communications systems, a  
QMSK protocol with its enhanced  spectral density of information encoding has the potential for significant impact. For instance,  a single $M=16$ QMSK channel offers an exponentially better HB as a function of the input energy in comparison to four multiplexed binary PSK channels (i.e. with equal encoding capacity), even though both arrangements consume an equal transmission band and energy per bit, see Supplementary Materials. This result shows that quantum measurement not only can improve the accuracy of a classical measurement, but also can offer new, heretofore unforeseen, advantages. Note that switching between the maximal energy efficiency mode and the spectrally efficient QMSK mode merely requires adjusting the parameters of the protocol, and it does not affect the hardware of either the transmitter or the receiver.

The quantum advantage accessed by our receiver can be applied to a broad range of measurements. Beyond classical communication, this method may be adopted to optimize coherent spectrometry with ultra-low signal (such as pump-probe spectrometry) \cite{Goun2016}. 

\begin{figure}
\centering
\includegraphics[width=0.3\textwidth]{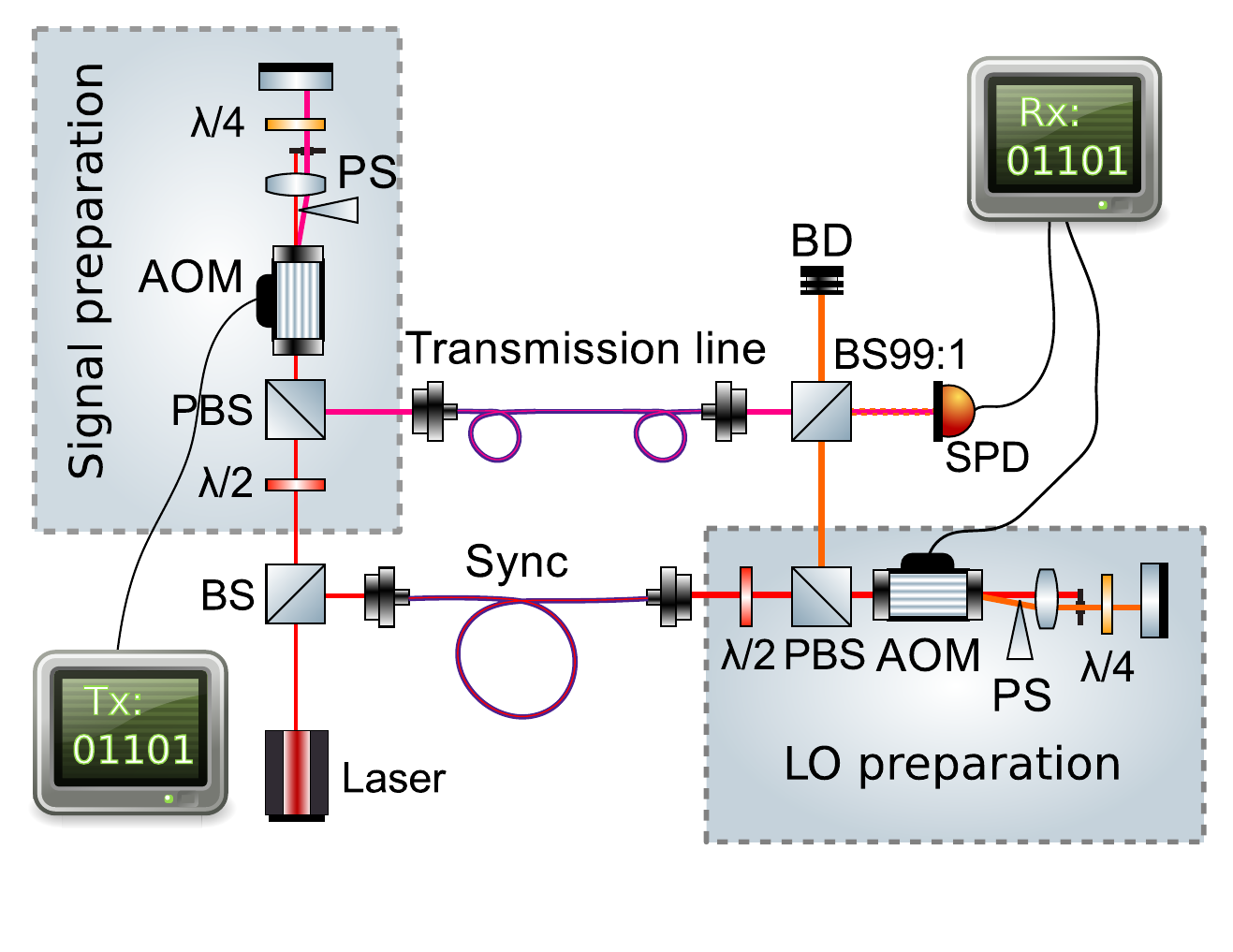}
\caption{\label{fig:exp} An experimental implementation of an $M$-ary CFSK receiver. The receiver discriminates between CFSK states prepared by a frequency-phase modulator implemented with a double-pass acousto-optical modulator (AOM). An identical AOM arrangement implements the adaptive local oscillator (LO). The displacement is performed by interfering the input state with the LO on a 99:1 beamsplitter (BS) followed by a single-photon detector (SPD). PBS - polarization beam splitter, PS - phase shifter, $\lambda/2$ and $\lambda/4$ half and quarter- wave-plates, BD - beam dump. 
}
\end{figure}

Despite the apparent complexity of CFSK, a proof-of-principle experimental implementation of the receiver is rather straightforward. A proposed encoding/decoding setup is based on a pair of double-pass acousto-optical modulators, and uses off-the-shelf components, see Fig. \ref{fig:exp}. The benefit of such a setup is its relatively low cost and a rapid path to success. Because the optical paths differ for different frequency detunings, the required phase correction can be applied with a simple wedge. This setup can be used to demonstrate alphabets with $\gtrsim 100$ symbols. In addition to the free-space implementation, the receiver can be implemented as an integrated design, where the local oscillator is modulated with a set of nested integrated Mach-Zehnder interferometers \cite{Izutsu} making it much more compatible with commercial applications.

\section {Conclusion}

In conclusion, we have introduced an optical energy efficient quantum receiver that 
experiences no degradation with increasing alphabet size. 
This receiver can be used to decrease the optical power required to transmit information by approximately three orders of magnitude as compared with the state-of-the-art commercial communication systems, while still efficiently using the channel resources. This advantage can also be used in deep space communication links to significantly enhance the power budget of satellites or increase the free-space communication range and communication rates. 
Quantum CFSK receivers can improve the amplification-free range by a factor of $\gtrsim$2 while using the existing global fiber infrastructure. To put this number in perspective, an $M$-ary CFSK receiver enables a fiber-based amplification-free 1 Gigabit/s communication link between Washington, DC and New York, NY powered by a $\approx$10 mW laser source, while a state-of-the art commercial receiver would require an input power over 10 W, which cannot be sent through a single fiber due to nonlinear effects and power damage. The quantum advantage offered by this receiver can also be used to further optimize spectral efficiency of communication channels. 

\section*{Funding Information}
OVT acknowledges partial support through the Russian Science Foundation grant No. 17-12-01079 and the Joint DFG-RFBR project CH 1591/2-1-16-52-12031 NNIO\_a.

\section*{Acknowledgements}
Authors thank Elizabeth A. Goldschmidt, Emanuel Knill, and Alan L. Migdall for fruitful discussion of the manuscript. 

\section*{Supplemental Documents}
See Supplement 1 for supporting content.



\bibliography{biblio}

\bibliographyfullrefs{biblio}


\end{document}